\documentstyle[psfig,aaspp4]{article}

\def\balpha{\hbox{$\alpha\hskip-7.2pt\alpha$}}
\def\bbeta{\hbox{$\beta\hskip-6.8pt\beta$}}
\def\etab{\hbox{$\eta\hskip-6.0pt\eta$}}
\def\bzeta{\hbox{$\zeta\hskip-5.8pt\zeta$}}
\def\vecr{{\rm \bf r}}

\begin{document}

\title{Superluminal Caustics of Close, Rapidly-Rotating Binary Microlenses}
\author{Zheng Zheng and Andrew Gould}
\affil{The Ohio State University, Columbus, OH 43210, USA}
\affil{E-mail: zhengz, gould@astronomy.ohio-state.edu}

\begin{abstract}

The two outer triangular caustics (regions of infinite magnification) 
of a close binary microlens move much faster than the components of the 
binary themselves, and can even exceed the speed of light.  When 
$\epsilon\ga 1$, where $\epsilon c$ is the caustic speed, the usual 
formalism for calculating the lens magnification breaks down. We develop 
a new formalism that makes use of the gravitational analog of the 
Li\'enard-Wiechert potential.  We find that as the binary speeds up, the 
caustics undergo several related changes: First, their position in space 
drifts.  Second, they rotate about their own axes so that they no longer 
have a cusp facing the binary center of mass.  Third, they grow larger 
and dramatically so for $\epsilon\gg 1$.  Fourth, they grow weaker roughly 
in proportion to their increasing size.  Superluminal caustic-crossing 
events are probably not uncommon, but they are difficult to observe.

\end{abstract}


\section{Introduction}

Microlensing is gravitational lensing where the separations between the 
images are too small to be resolved. The directly detectable consequence 
of microlensing is that the brightness of the source varies in a way 
determined by the lens properties and the projected lens-source trajectory. 
Paczy\'nski (1986) pointed out that microlensing would be a useful tool 
to detect massive compact halo objects. Microlensing surveys have since 
been carried out towards the Galactic bulge, the Magellanic Clouds, and 
M31. About 500 microlensing events have been detected to date (see Mao 2000
for a review).

Gravitational lensing by two point masses was carefully studied by 
Schneider \& Weiss (1986). The most striking feature of such binary 
lensing is its caustics - one or several closed curves in the source 
(objects to be imaged) plane where a point source is infinitely magnified 
by the lens. As a reflection of the caustic structure of the magnification, 
the light curves of such binary lensing events may have multiple peaks. 
Microlensing surveys have detected about 30 such events (e.g., Udalski et 
al.\ 1994; Alcock et al.\ 2000). Star-planetary systems are an extreme 
form of binary. Mao \& Paczy\'nski (1991) first suggested that extrasolar 
planetary systems could be discovered by microlensing surveys.

Dominik (2000) conducted the first systematic investigation of the effect 
of binary rotation on microlensing light curves.  Although all physical 
binaries rotate, static models suffice to reproduce the light curves of 
the great majority of observed microlensing events, even those with superb 
data such as MACHO 97-BLG-28 (Albrow et al.\ 1999a) and MACHO 98-SMC-1 
(Afonso et al.\ 2000).  The only event observed to date for which a 
rotating model is {\it required} is MACHO 97-BLG-41 (Albrow et al.\ 2000), 
and static models are excluded for this event only because the source
traverses two disjoint caustics, a rare (so far, unique) occurrence.
Bennett et al.\ (1999) had earlier proposed that the odd light curve of
MACHO 97-BLG-41 could be explained by a triple-lens system consisting of a 
binary plus a jovian-mass planet.

In the treatments of binary rotation given to date (Dominik 1998; Albrow 
et al.\ 2000), the light curve is actually calculated by considering a 
series of {\it static} binaries, each with the configuration of the binary 
being modeled at the {\it instant} when the light ray from the source 
passes the plane of the center of mass of the lens.  That is, the deflection 
of light by the binary, $\balpha$, is taken to be the vector sum of the 
deflections produced by the two components of the binary, 
$\balpha=\balpha_1+\balpha_2$, according to the Einstein (1936) formula,
$$
\balpha_i = -{4 G M_i\over b_i^2c^2}{\bf b}_i. \eqno(1) 
$$
Here $M_i$ is the mass, and ${\bf b}_i$ is the impact parameter of the
$i$th component of the binary.

This approach is strictly valid only in the limit
$$
\epsilon \ll 1,\qquad \epsilon \equiv {\omega b\over c}, \eqno(2)
$$
where $2\pi/\omega$ is the period of the binary.  For MACHO 97-BLG-41, 
the only microlensing event for which rotation has been measured
(Albrow et al.\ 2000), $\epsilon\sim 10^{-4}$, so this approach is
certainly valid.  However, in principle $\epsilon$ can be close to unity
or can even greatly exceed unity.  In this case, it is necessary to take 
account of the binary motion during the time that the source light is 
passing close $(\la b)$ to the lens plane.  The Einstein formula (1), 
which was calculated for a static lens, is then no longer valid.

Here we study the rotation effects on the caustic structure of close,
rapidly-rotating binary lenses. We present our main idea and method in 
\S\ 2 and discuss the rotation effects in \S\ 3. In \S\ 4, we summarize 
our results and discuss some possible applications.

\section{Main Idea and Method}

As mentioned in \S\ 1, the binary phase varies during the time that the 
photons are traveling from the source to the observer.  This modifies the 
calculation of the instantaneous magnification map, especially for 
$\epsilon\ga 1$.  The retarded gravitational potential then begins to 
differ significantly from the naive Newtonian potential, which would 
normally be adequate in the weak-field limit and which is used to derive 
equation (1).

\subsection{Retarded Gravitational Potential}

As an approximate result of Einstein's field equations, the deflection of 
a light ray passing through a static gravitational field can be expressed 
as an integral of the gradient of the Newtonian gravitational potential 
performed along the trajectory of the light (Bourassa et al.\ 1973):
$$
\balpha=
-{2 \over c} \int^{+\infty}_{-\infty} \nabla\phi{\rm d}t, \eqno(3)
$$  
which yields equation(1). However, for the non-static case, the 
configuration of the gravitational field will propagate at light speed, 
and we must instead use the retarded potential.

In analogy to the results of classical electrodynamics (e.g. Jackson 1975),
the gravitational potential at field point \vecr\ and time $t$ is 
contributed by every mass point \vecr$^\prime$ at an earlier time 
$t^\prime=t-|\vecr-\vecr^\prime|/c$, 
$$
\phi(\vecr,t)=-\int{G\rho(\vecr^\prime,t^\prime)
\over |\vecr-\vecr^\prime|}{\rm d^3r^\prime}, \eqno(4)
$$
where $\rho$ is the (time-dependent) mass distribution. 
  
For a point mass $M$, the retarded potential (4) can be written, similarly
to the Li\'enard-Wiechert potential,
$$
\phi(\vecr,t)=-{GM \over (1-{\bf n^\prime 
\cdot \bbeta^\prime})|\vecr-\vecr^\prime|},           \eqno(5)
$$
where ${\bf n^\prime}=(\vecr-\vecr^\prime)/|\vecr-\vecr^\prime|$, 
$\bbeta^\prime$ is the velocity of the point mass divided by $c$, and the 
prime denotes the value at time $t^\prime$,
$$
t=t^\prime+{|\vecr-\vecr^\prime(t^\prime)| \over c}. \eqno(6)
$$

The Newtonian gravitational field, ${\rm \bf g}=-\nabla\phi$, is then
$$
{\bf g}(\vecr,t)=
\biggl[\nabla{GM \over (1-{\bf n \cdot \bbeta})|\vecr-\vecr^\prime|}
\biggr]_{t^\prime} 
=\biggl[-{GM \over (1-{\bf n \cdot \bbeta})^2|\vecr-\vecr^\prime|^2}({\bf n}+
\bbeta)\biggr]_{t^\prime}. \eqno(7)
$$

For those rays that pass across the lens at a distance much greater than 
the Schwarzschild radius $2GM/c^2$, the total deflection angle caused by 
several lens objects is a superposition of the individual deflections 
(Bourassa et al.\ 1973). From equations (3) and (7), we have,
$$
\balpha=-\sum_i{2GM_i \over c}
\int^{+\infty}_{-\infty}\biggl[{ {\bf n}_i+\bbeta_i \over 
(1-{\bf n_{\it i} 
\cdot \bbeta_{\it i}})^2|\vecr-\vecr_i|^2}\biggr]_{t^\prime}{\rm d}t. \eqno(8)
$$

It is convenient to take the time when the photon crosses the lens plane as 
$t = 0$. Then at a distance $|ct|$ to the lens plane, the photon will feel 
the potential caused by the point mass $M_i$ at time $t^\prime$. We have
$$
|\vecr-\vecr_i(t^\prime)|=\sqrt{b_i^2(t^\prime)+(ct)^2}, \eqno{(9)}
$$ 
where $b_i(t^\prime)$ is the distance from the point mass $M_i$ to the 
impact point at time $t^\prime$. Substitution of equation (9) into equation
(6) yields
$$
t={1 \over 2}t^\prime-{b_i^2(t^\prime) \over 2c^2t^\prime}. \eqno(10)
$$

This relation between $t$ and $t^\prime$ reflects the retarded effect. For
a finite $b_i(t^\prime)$, and under the condition $t>t^\prime$, we find 
that when $t \rightarrow -\infty$, $t^\prime \rightarrow - \infty$ and 
when $t \rightarrow + \infty$, $t^\prime \rightarrow 0^-$. That is 
$t^\prime \simeq 2t$ when $t \ll 0$ and $t^\prime \simeq 0$ when $t \gg 0$. 
For a circular orbit, this means that as a photon moves towards the 
rotating system, it will ``find'' that the angular velocity of the system 
is nearly doubled, and as it moves away from the system, it will feel an 
almost static field. Equation (10) makes it possible to replace the 
integration variable in equation (8) with $t^\prime$ (from $-\infty$ to $0$).

\subsection{Lens Equation}

The lens equation, i.e.\ the light-ray deflection equation, tells one how 
the light-ray deflection maps points in the source plane into points in the 
image plane (lens plane). If a photon comes from point $\etab$ in the 
source plane and hits point $\bzeta$ in the lens plane, we have (Schneider 
\& Weiss 1986),
$$
\etab={D_s \over D_l}\bzeta-D_{ls}\balpha(\bzeta), \eqno(11)
$$
where $D_s$ and $D_l$ are the distances from the observer to the source 
plane and to the lens plane, respectively, and $D_{ls}=D_s-D_l$ is the 
distance between the source and the lens. 

In this paper, we consider a simple case: the lens is composed of two
stars with equal masses $M_1=M_2$, rotating about each other in a 
circular face-on orbit. We define the distance between two stars to be 
$2a$ and the angular velocity around the center of mass to be $\omega$ 
(see Fig.~1). We define the radius of the Einstein ring generated by the 
total mass $M_1+M_2$ to be (e.g. Schneider \& Weiss 1986) 
$$ 
r_{\rm E}=\sqrt{{4G(M_1+M_2) \over c^2}{D_lD_{ls} \over D_s}}, \eqno(12)
$$
and then normalize the coordinates of points at the lens plane and 
those at the source plane using this radius and the radius projected onto
the source plane, respectively:
$$
X={a \over r_{\rm E}}, \eqno(13)
$$
$$ 
{\bf r}={\bzeta \over r_{\rm E}}, \eqno(14)
$$
$$
{\bf x}=\etab  \biggl[r_{\rm E} {D_s \over D_l}\biggr]^{-1}. \eqno(15)
$$
The consequent lens equation is given in Appendix A.
 
\begin{figure}
\centerline{\psfig{figure=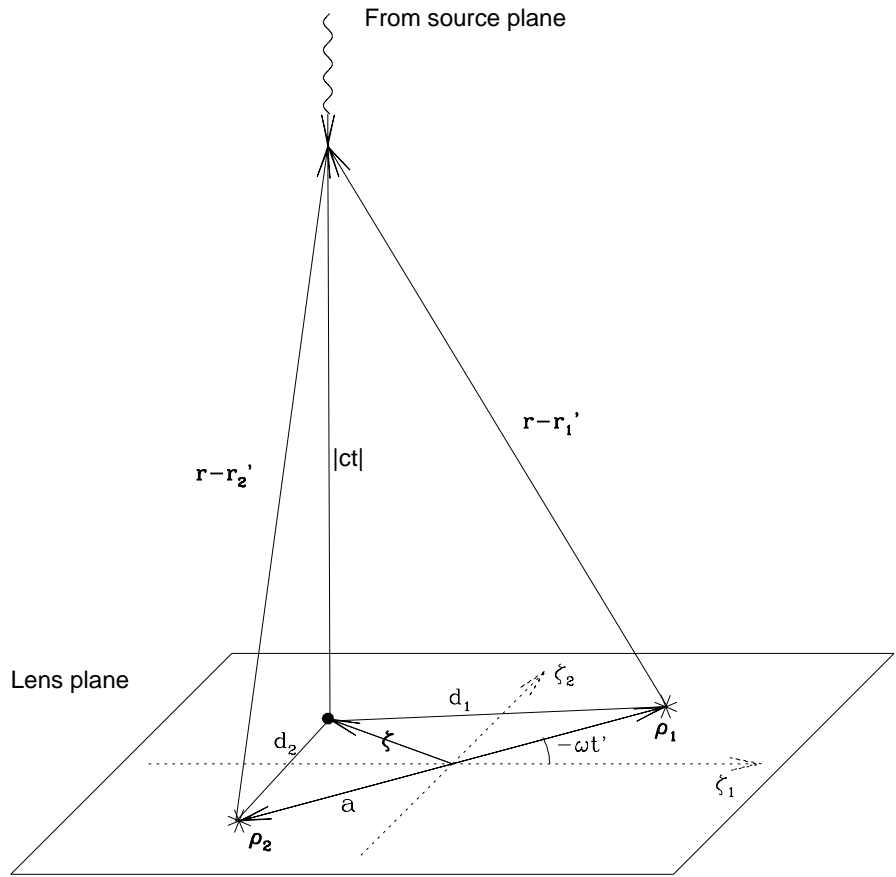,width=14cm,height=14cm}}
\caption[]{Geometry of the rotating binary lens with equal masses 
and face-on circular orbit. At time $t$, the photon feels a potential 
generated by the two stars at an earlier time $t^\prime$.}
\end{figure}

The focus of the present study is the case $X\ll 1$, for which there are 
two small triangular caustics lying at projected distances
$\sim r_{\rm E}/(2 X)$ from the binary center of mass (Schneider \& Weiss 
1986). As the binary rotates at angular speed $\omega$, the caustics 
rotate with it.  Hence the transverse speed of the caustics is
$v/(2X^2)$, where $v=\omega a$ is the speed of the binary components.
For very small $X$, the caustics can move much faster than the speed of
light (``superluminal motion'') even when the binary itself is well
within the non-relativistic regime.

Since we have fixed the mass ratio, $M_2/M_1=1$, and adopted a face-on, 
circular orbit, there are only three free parameters of the lens system, 
$X$, $\omega$, and $a$.  However, from the standpoint of studying the 
caustic structure that appears in diagrams from which all physical
dimensions have been scaled out, it is only necessary to consider
{\it dimensionless} parameters.  There are two independent such parameters.
One is $X$.  There are two obvious possible choices for the other,
$$
\beta = {\omega a\over c},\qquad 
\epsilon = {\omega r_{\rm E}\over 2 c X} = {\beta \over 2 X^2}.  \eqno(16)
$$
The first is the speed of the lenses as a fraction of the speed of light,
and the second is approximately the speed of the caustics as a fraction of 
the speed of light. As we show in the next section, $\epsilon$ is a more 
useful parameter than $\beta$ because the caustics are more directly 
affected by their own speed rather than that of the lenses.

To make a two-dimensional magnification map of the lens system (around the
caustics), we use the inverse ray-shooting technique (e.g., Schneider \&
Weiss 1986; Wambsganss 1997). Uniformly distributed light rays in the lens 
plane are evolved back to the source plane according to the lens equation. 
The magnification of each point in the source plane is then proportional to 
the density of rays at this point. We study the cases of $X=0.1$, $X=0.05$ 
with various values of $\epsilon$. 
    
\section{Effects of the Rotation}

Figure 2 displays some examples of the outer caustics generated by adopting 
different parameters. Compared with the static case, these caustics show 
some new features.
 
\subsection{``Orbit Position'' of the caustic}

In the static case, the outer caustics of an equal-mass binary are located 
along the perpendicular bisector of the binary. If the binary lens rotates 
around the center while the photon is traveling, it is not hard to imagine 
that the consequent caustics will drift with respect to the bisector of the 
binary lens at phase $t=0$. It turns out that the direction of this drift 
is opposite to the rotation. Physically, the reason for this ``opposite'' 
drifting is that at {\it all} times $t$, the phase of the binary 
corresponds to an earlier time $t^\prime<0$. See equation (10) and the 
analysis following it. In Figure 3, we show the angular position of the 
caustic relative to the static case as a function of $\beta$. For 
$\beta \la 0.1$, we find a fitted formula for the ``orbit'' angle 
$\theta_{orbit}$ (in radians):
$$
\theta_{orbit}={4 \over 7}\beta={8 \over 7}X^2\epsilon. \eqno(17)
$$
We find that the fitted coefficient in equation (17) is very close to 
the ratio of two small integers (4/7), but we do not know whether this
result is exact.
  
\begin{figure}
\centerline{\psfig{figure=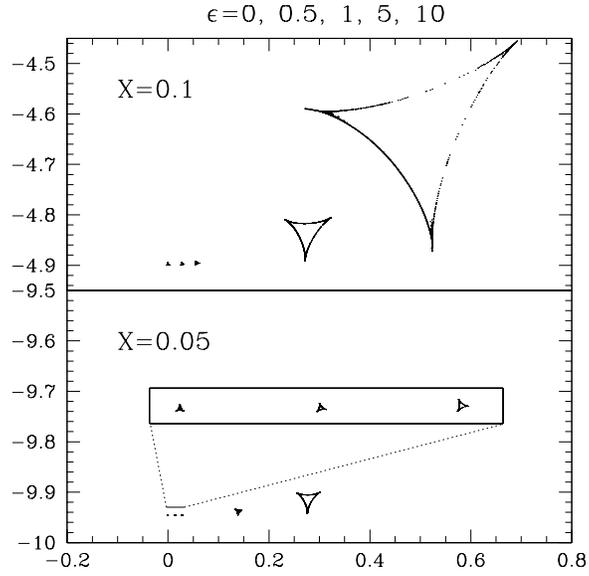,width=8cm,height=8cm}
           }
\caption[]{The outer caustics of a rapidly rotating binary lens.
The upper and lower panels show the $X=0.1$ and $X=0.05$ cases, respectively.
In each panel, caustics are shown $\epsilon=0$, 0.5, 1, 5, 10, from 
left to right, respectively. In the lower panel, the three smallest
caustics are also shown expanded by a factor 20.}
\end{figure}

\begin{figure}
\centerline{\psfig{figure=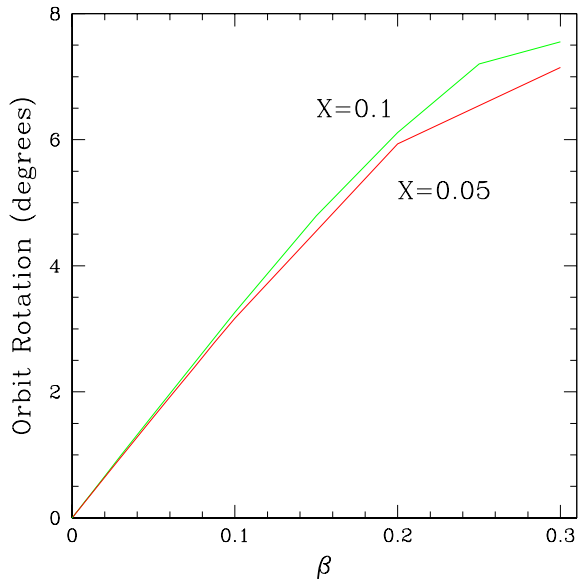,width=8cm,height=8cm}
           }
\caption[]{The orbit rotation angles of the outer caustics as a 
function of $\beta$.}
\end{figure}

\subsection{``Spin'' - Pointing of the caustic}
Apart from the outer caustics' orbit motion as a whole, these 
``concave triangles'' have their own rotation. We use the direction of a 
vertex (the one pointing to the center of mass in the static case) as a 
tracer for the caustic's ``spin''. Unlike the Moon which always shows 
the same hemisphere to the Earth, these triangles spin much faster than 
their orbital rotation. For the non-static case, they will no longer point 
to the center of mass. See Figure 4. The fitted formula for the spin
angle $\theta_{spin}$ (in degrees) for $\epsilon <2$ is
$$
\log\theta_{spin}=31.32\,\log\epsilon+0.02154. \eqno(18)  
$$

\begin{figure}
\centerline{\psfig{figure=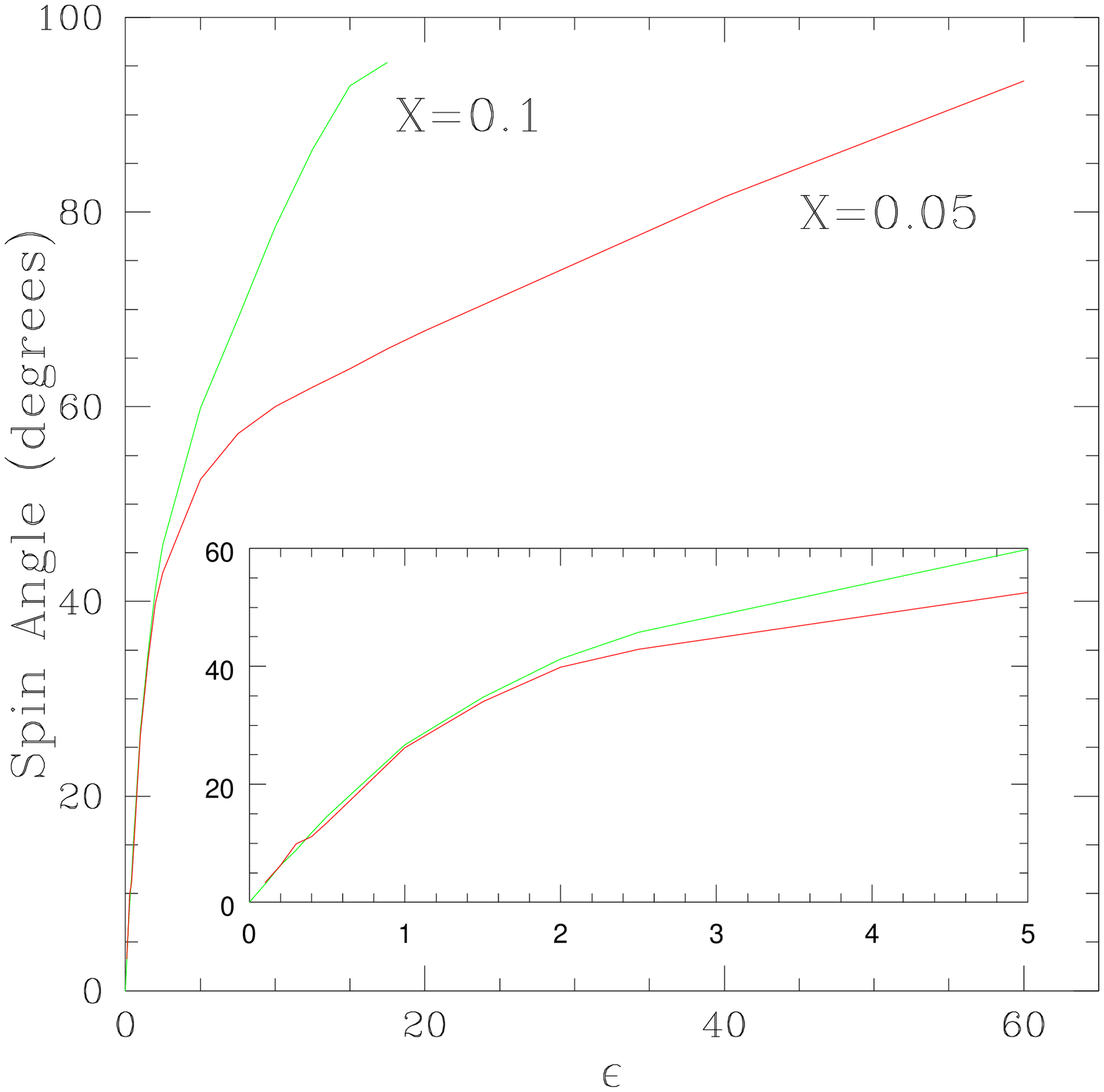,width=8cm,height=8cm}
           }
\caption[]{The spin angle of the outer caustics as a function of $\epsilon$.
The inset details the behavior for $\epsilon < 5$.}
\end{figure}

\subsection{``Expansion'' - Enlargement of the caustic} 

For static case, it is known that the closer the binary is, the farther
the triangular caustics are away from the center of mass and the smaller 
they become. The outer caustics shrink almost to a point in the case of 
very small $X$. However, after taking the rotation effect into account, 
we find that the tiny caustics are strikingly magnified. Meanwhile, unlike 
the static case, the shape of the caustics gradually loses its symmetry. We 
choose the area inside the caustic as a measure of the expansion effect. 
Since in the static case, the linear size of the outer caustic scales 
approximately as $X^3$, we normalize the area in our case by $X^6$. The 
expansion effect is illustrated in Figure 5.

\begin{figure}
\centerline{\psfig{figure=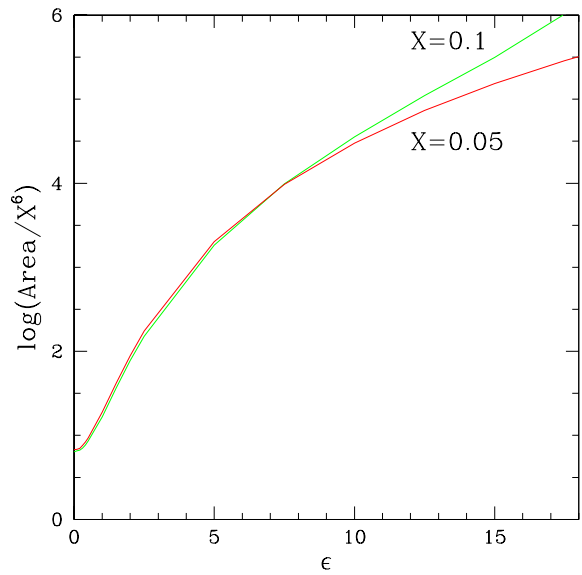,width=8cm,height=8cm}
           }
\caption[]{The expansion effect of the outer caustic. The area inside the 
caustic is normalized by $X^6$.}
\end{figure}

\subsection{Magnification Properties}

The rapidly rotating binary lens makes the outer caustics move with a
speed comparable to light speed which in turn brings some new phenomena
to the magnification properties of the caustics. 

First, the expansion of the outer caustic dilutes the magnification.
In the static case, the magnification factor near a caustic curve can be
described as $A(u)=A_0+(\Delta u_\perp /u_r)^{-1/2}$, where $\Delta u_\perp$ 
is the perpendicular distance to the caustic and $A_0$ and $u_r$ are 
constants (Schneider \& Weiss 1986; Schneider \& Weiss 1987; Albrow et al.\ 
1999b). Hence $u_r$ describes the strength of the caustic. We investigate 
the variation of $u_r$ as a function of $\epsilon$ and find that rapid 
rotation weakens the strength of the caustics. The relative strength of 
the three caustic lines (of the triangular caustic) also changes (see 
Fig.~6 for the case $X=0.1$): one caustic line (left border, see Fig.~2) 
becomes the strongest one at large $\epsilon$ (large area). With a 
dimension of linear size, $u_r$ is expected to scale as $X^3$. However, our 
calculation shows that there is a slight deviation from this scale law: 
the change of $u_r$ for $X=0.05$ is a little steeper than that for 
$X=0.1$. Taking account of our resolution, we are not sure whether this 
marginal effect is real or not.      

Secondly, the velocity of the source with respect to the outer caustic is
overwhelmingly determined by the caustic's high speed. The relative 
source-caustic trajectory is then a small piece of arc centered at the 
binary center of mass. The rapid rotation implies that the timescale for 
crossing the caustic will be very short. We choose the square root of the 
area $S$ inside the caustic as the size of the caustic and the crossing 
time is then
$$
t_c={\sqrt{S} \over \epsilon}{r_{\rm E} \over c},   \eqno(19)
$$
where $r_{\rm E}$ is the Einstein ring defined in equation (12). Since the 
area scales very nearly as $X^6$ (see Fig 5), this time scale can be 
normalized by $X^3$. As shown in Figure 7, $t_c$ has a minimum near 
$\epsilon \sim 1$. At smaller $\epsilon$, $t_c$ becomes larger mainly due 
to the ``low'' speed of the caustic and at larger $\epsilon$, mainly due
to the expansion of the area.    

\begin{figure}
\centerline{\psfig{figure=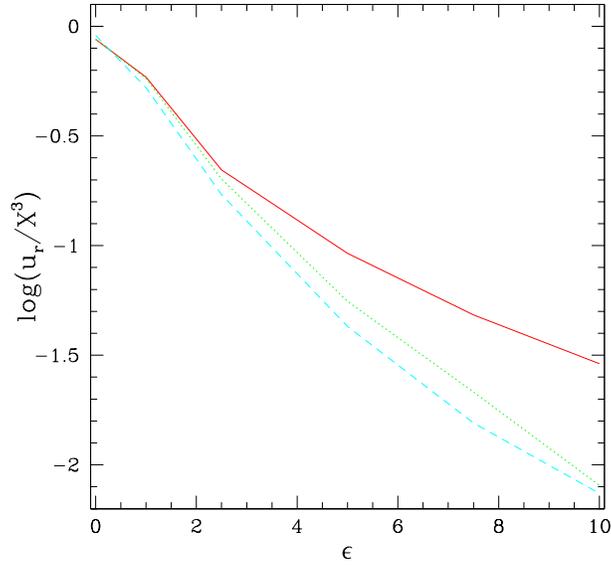,width=8cm,height=8cm}
           }
\caption[]{The strength $u_r$ of the three caustic lines of the outer 
triangular caustic as a function of $\epsilon$ (for the case $X=0.1$, 
normalized by $X^3$). See text for the definition of $u_r$. The 
counterparts of the left, bottom and right caustic lines in the static 
case (see Fig.~2) are represented by the solid line, dashed line and 
dotted line, respectively.}
\end{figure}

\begin{figure}
\centerline{\psfig{figure=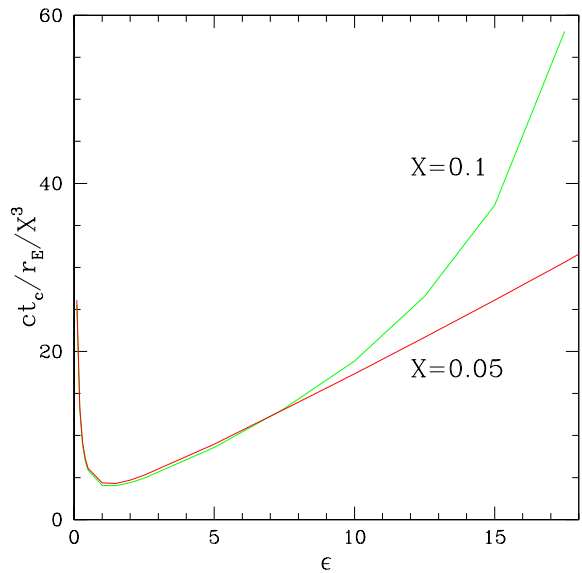,width=8cm,height=8cm}
           }
\caption[]{Time scale to cross the outer caustics in units of $r_{\rm E}/c$,
normalized by $X^3$, as a function of $\epsilon$.}
\end{figure}

\section{Discussions}

In this paper, we point out that a modification is necessary to get the 
instantaneous magnification map for close, rapidly-rotating binary lenses, 
in which case the outer caustics have a very high speed which can even be 
superluminal. Taking the retarded gravitational potential into 
consideration, we investigate the outer caustic behavior for such a lens 
with equal masses and face-on circular orbit. Compared with the static 
case, the caustic is displaced in orbit position and is rotated about 
its own axis. The most remarkable result is the enlargement of the caustic 
by the rapid motion of the  lens. This increase in size induces a 
corresponding drop in the strength of the caustic.

Instead of starting from Einstein's field equation, we use a retarded 
potential at the first step. Although strictly speaking this method has 
its limitations, it is a reasonable approach for our purpose since our 
analysis focuses on the high speed of the outer caustics while 
the binary itself is not in the extreme relativistic regime.  

What would be required to observe superluminal caustics?  Combining
the definitions of $\epsilon$ and $r_{\rm E}$ with Kepler's Third Law 
yields,
$$
\epsilon = {(2 G M)^{3/2} D\over c^3 a^{5/2}} =
0.37
\biggl({M\over M_\odot}\biggr)^{3/2}
\biggl({a\over 0.1\,\rm AU}\biggr)^{-5/2}
{D\over 2\,\rm kpc},                       \eqno(20)
$$
where $D=D_l D_{ls}/D_s$, $M$ is the mass of one binary component, and
where we have normalized to the case of a pair of solar-mass stars seen
half way to the Galactic center.  Equation (20) can be rewritten in terms
of
$X$,
$$
X = 2^{-9/10}\epsilon^{-2/5}\biggl({G M\over D c^2}\biggr)^{1/10}. \eqno(21)
$$
Hence, to obtain $\epsilon\sim 1$ would require $X\sim 0.012$, a factor 4 
smaller than even the lesser of the two values that we examined in this 
paper.  Note that this result is extremely insensitive to either $M$ or $D$.

From Figure 5, the combined cross section (linear size) of the
two caustics at $\epsilon\sim 1$ is $S^{1/2}\sim 8 X^3 r_{\rm E}$, that is, 
a factor $4 X^3$ smaller than for the lens itself.  For $X=0.012$, this 
factor is $10^{-5}$, so that at first sight it appears completely hopeless 
that superluminal caustics would ever be observed.  However, the event rate
is the product of the cross section with the transverse speed, and the
caustic moves $\sim 10^3$ times faster than transverse speed of the binary
center of mass.  In fact, since the caustic is likely to be smaller than the
source, the event rate is given by the source size times the transverse
speed of the caustic.  Thus, the ratio of superluminal-caustic events to
normal events generated by the same binary is
$$
{\Gamma_{\rm super}\over \Gamma_{\rm normal}}
\sim {2R_* D_l/D_s\over r_{\rm E}}\, {\epsilon c \over v_\perp}
\sim 1.2\,{R_*/R_\odot\over r_{\rm E}/6\,\rm AU}\, {\epsilon\over
v_\perp/200\,\rm km\,s^{-1}},                 \eqno(22)
$$
where $R_*$ is the radius of the source, and $v_\perp$ is the transverse
speed of the binary center of mass. That is, they are about equally likely.
Note that only a small minority (roughly a fraction $X$) of the 
superluminal events occur in association with a normal event (where 
the source passes within the Einstein ring).  The rest are isolated 
``spike events''.  The real problem with observing superluminial events 
is not that they are uncommon, but that they are weak. For $\epsilon\la 1$ 
the caustic covers (and hence magnifies) only a small fraction of the 
source star.  For example, for $X=0.012$, $M=M_\odot$, $D_l=D_{ls}=4\,$kpc, 
and $R=R_\odot$, the caustic covers only 0.01\% of the source.  Of course, 
for higher $\epsilon$, the caustic area grows, but as we discuss in 
\S\ 3.4, the strength of the caustic declines. Hence, at least for the 
present, superluminal caustics appear to be of mainly theoretical interest.
        
\acknowledgments{We thank Scott Gaudi for valuable discussions.
This work was supported in part by grant AST 97-27520 from the NSF.}

\centerline{\bf APPENDIX A}

The lens equation can be written in terms of the normalized coordinates 
of points at the lens plane ${\bf r}=(r_1, r_2)$ and those at the source 
plane ${\bf x}=(x_1,x_2)$. It is convenient for calculation if we replace 
the integration variable $t$ in equation (8) with $t^\prime$ using equation
(10). Note that $-\infty < t^\prime <0$. According to the geometric
relations in Figure 1, with the definition $\tau=ct^\prime / a$, we then 
have the lens equation in component form:
$$
x_i=
r_i+{1 \over 4} \int^{0}_{-\infty}[F_{1i}(\tau)+F_{2i}(\tau)]X{\rm d}\tau,
\eqno(A1)
$$ 
where $i=1,2$ and
$$
F_{i1}=
{4X\tau[r_1+(-1)^iX\cos\beta\tau] \over (1-p_i)[(X\tau)^2+d_i^2(\tau)]^2}
-
{2(-1)^i\beta\sin\beta\tau \over (1-p_i)[(X\tau)^2+d_i^2(\tau)],}
\eqno(A2)
$$
$$
F_{i2}=
{4X\tau[r_2-(-1)^iX\sin\beta\tau] \over (1-p_i)[(X\tau)^2+d_i^2(\tau)]^2}
-
{2(-1)^i\beta\cos\beta\tau \over (1-p_i)[(X\tau)^2+d_i^2(\tau)],}
\eqno(A3)
$$
$$
d_i^2(\tau)=[r_1+(-1)^iX\cos\beta\tau]^2+[r_2-(-1)^iX\sin\beta\tau]^2,
\eqno(A4)
$$
$$
p_i=(-1)^{i-1}{2X\beta\tau \over (X\tau)^2+d_i^2(\tau)}
(r_1\sin\beta\tau+r_2\cos\beta\tau).
\eqno(A5)
$$


\end{document}